\documentclass[prl,aps,amssymb,showpacs,twocolumn,superscriptaddress,times,10pt]{revtex4-1}

\usepackage[english]{babel}
\usepackage[pdftex,colorlinks=false]{hyperref}
\usepackage{amsmath,amssymb,amsthm}

\usepackage{amsmath,amssymb,mathrsfs}
\usepackage{latexsym}
\usepackage{graphicx}
\usepackage{epstopdf}
\usepackage{color}
\usepackage{amsfonts}
\usepackage{bm}
\usepackage{bbm}
\usepackage{multirow}
\usepackage{subfiles}

\usepackage{natbib}

\usepackage{ulem} 

\usepackage{tikz} 
\usepackage{tikz-feynman}

\usepackage{mathtools} 
\usepackage{physics}
\usepackage[version=3]{mhchem}

\usepackage{microtype} 
\usepackage{csquotes} 
\usepackage{booktabs} 

\makeatletter
\renewcommand\@make@capt@title[2]{%
	\@ifx@empty\float@link{\@firstofone}{\expandafter\href\expandafter{\float@link}}%
	{\textbf{#1}}\@caption@fignum@sep#2\quad
}%
\makeatother

\makeatletter
\g@addto@macro\bfseries{\boldmath} 
\makeatother

\usepackage{graphicx}
\usepackage{placeins}

\usepackage{siunitx} 
\DeclareSIUnit\year{a}
\DeclareSIUnit\pixel{px}
\DeclareSIUnit\line{line}

\usepackage{scrlayer-scrpage} 
\usepackage{lastpage} 

\newcommand{\orb}[2]{\(\text{#1}_\text{#2}\)}

\newcommand{\cre}{{\dag}}
\newcommand{\ann}{{\vphantom{\dag}}}
\newcommand{\Tc}{$T_c$}

\newcommand{\mcomma}{\,\text{,}}

\newcommand\ddfrac[2]{\frac{\displaystyle #1}{\displaystyle #2}} 

\graphicspath{./figures/}

\def\ie{\textit{i.e.},\ }

\def\ea{\textit{et al.}}
\newcommand{\bs}[1]{\boldsymbol{#1}}

\begin{document}

	\title{From high-\Tc{} to low-\Tc{}: Multi-orbital effects in
          transition metal oxides}

     \author{Michael Klett}
     	 \thanks{These two authors contributed equally }
     \affiliation{Institute for Theoretical Physics, University of Wuerzburg, D-97074 Wuerzburg, Germany}
     \author{Tilman Schwemmer}
     	\thanks{These two authors contributed equally }
     \affiliation{Institute for Theoretical Physics, University of Wuerzburg, D-97074 Wuerzburg, Germany}

	\author{Sebastian Wolf}
	\affiliation{School of Physics, University of Melbourne, Parkville, VIC 3010, Australia}

	\author{Xianxin Wu}
	\affiliation{Max-Planck-Institut f\"ur Festk\"orperforschung, Heisenbergstrasse 1, D-70569 Stuttgart, Germany}
	\affiliation{ Beijing National Laboratory for Condensed Matter Physics,}

	\author{David Riegler}
	 \affiliation{Institute for Theoretical Physics, University of Wuerzburg, D-97074 Wuerzburg, Germany}

	\author{Andreas Dittmaier}
	\affiliation{Institute for Theoretical Physics, University of Wuerzburg, D-97074 Wuerzburg, Germany}

	\author{Domenico Di Sante}
	\affiliation{Institute for Theoretical Physics, University of Wuerzburg, D-97074 Wuerzburg, Germany}

	\author{Gang Li}
	\affiliation{School of Physical Science and Technology, ShanghaiTech University, Shanghai 201210, China}
	\affiliation{ShanghaiTech Laboratory for Topological Physics, ShanghaiTech University, Shanghai 201210, China}

	\author{Werner Hanke}
	\affiliation{Institute for Theoretical Physics, University of Wuerzburg, D-97074 Wuerzburg, Germany}

	\author{Stephan Rachel}
	\email{stephan.rachel@unimelb.edu.au}
	\affiliation{School of Physics, University of Melbourne, Parkville, VIC 3010, Australia}

	\author{Ronny Thomale}
	\email{rthomale@physik.uni-wuerzburg.de}
	\affiliation{Institute for Theoretical Physics, University of Wuerzburg, D-97074 Wuerzburg, Germany}

\begin{abstract}
Despite the structural resemblance of certain cuprate and
nickelate parent compounds there is a striking spread of \Tc{} among such transition
metal oxide superconductors. We adopt a minimal two-orbital $e_g$
model which covers
cuprates and nickelate heterostructures
in different parametric limits, and analyse its superconducting
instabilities. The joint consideration of interactions, doping, Fermiology,
and in particular the $e_g$ orbital splitting allows us to explain the
strongly differing pairing propensities in cuprate and nickelate superconductors.
\end{abstract}

\date{\today}
\maketitle


\textit{Introduction.}---High-temperature unconventional superconductivity, as discovered in
copper oxides in 1986~\cite{bednorz}, has ever since decisively framed the landscape of
research in condensed matter physics. In particular, many trends of \Tc{} as a function of tunable system
parameters have been investigated. The hope is to identify a way to
tune the cuprates just enough to approach room temperature, and hence
render them, or another class of unconventional
superconductors, technologically viable~\cite{keimer,RevModPhys.84.1383}.
Initially, the rather
accurate single-orbital one-band Hubbard model description was
assumed to be an advantageous, and hence already \Tc{} optimized, feature of
the copper oxide superconductors. Due to the Jahn-Teller effect, the
apical oxygen distance to the CuO$_2$ planes is elongated, and thus
ensures a splitting of the $e_g$ orbitals such that the $3d^9$
configuration of Cu yields a nearly exclusive hole population of the
$3d_{x^2-y^2}$ orbital~\cite{pavarini_band-structure_2001}.

Research on unconventional superconductivity
over the past two decades indicates that this picture, in its generality, needs
revision \cite{PhysRevLett.105.057003}. While orbital fluctuations may be
detrimental to high-\Tc{},
multi-orbital systems can also yield
beneficial effects for unconventional superconductivity, such as the
multi-pocket Fermiology of iron-based
superconductors~\cite{si,platt_functional_2013}.
Interpreting the single-band cuprates as the nucleus for high-\Tc{}
superconductivity, the addition of multi-orbital character can thus take
different turns.
Nickelate thin film heterostructures such
as \ce{LaNiO3}/\ce{LaAlO3}, whose electronic structure is suggested to be
analogous to the
high-\Tc{} cuprate compounds~\cite{chaloupka_orbital_2008,hansmann_turning_2009},
only show a \Tc{} as low as 3 K~\cite{lanio3hetero}.
This might hint at the
detrimental effect of orbital fluctuations, which would be in line with a precise analysis of the impact of interactions on the orbital
polarization~\cite{PhysRevLett.107.206804}. We call this a low-\Tc{} instance of multi-orbital effects.
To the contrary, \ce{Ba2CuO_{3+\delta}}, where $e_g$ orbital fluctuations
are likewise expected to be as prominently present as in LaNiO$_3$
heterostructures, reaches a \Tc{} as high as 70 K~\cite{Li12156},
which we highlight as a high-\Tc{} instance.
Note that all aforementioned
material examples of transition metal oxide superconductors are
characterized by the $e_g$ orbitals of the transition metal atom at
low energies,
albeit for different orbital fillings (Ni $3d^7$ versus Cu $3d^9$), doping
levels, and Fermi surface topologies (Fermiologies).
Adopting the view from a generalized $e_g$ two-orbital octahedral oxide
setting, Ba$_2$CuO$_{3+\delta}$ has
recently been speculated~\cite{Li12156} to drastically modify the $e_g$
splitting due to a strong reduction of the apical oxygen distance, bringing into
play both the $3d_{x^2-y^2}$ and $3d_{3z^2-r^2}$ orbital.

In this Letter, we particularize on the analysis of an effective two-band model spanned by the $e_g$ orbital space of transition metal oxides, and investigate the onset of superconducting order.
While variants of this model have already been studied in the context of overdoped cuprate superconductors in general \cite{jiang_nodeless_2018}
and \ce{Ba2CuO_{3+\delta}} in particular \cite{maier_d_wave_2018},
we create a synoptic perspective on multi-orbital effects by comparing the high-\Tc{} material \ce{Ba2CuO_{3+\delta}} (BCO)
to the low-\Tc{} regime of \ce{LaNiO3}/\ce{LaAlO3} (LNO/LAO) heterostructures.
The rare-earth nickelates, with a close similarity between the \ce{NiO2} and \ce{CuO2} planes, have recently surfaced as potentially cuprate related unconventional superconductors \cite{PhysRevB.101.060504} and, since the finding of \Tc{} up to 15 K in \ce{NdNiO2}, have established an exciting domain, in which even higher \Tc's may be realized \cite{natureHwang}.
An earlier idea in this direction \cite{chaloupka_orbital_2008},
followed up by LDA+DMFT electronic structure calculations \cite{hansmann_turning_2009},
suggested turning a nickelate Fermi Surface (FS) into a cuprate-like one by orbital engineering via heterostructuring:
sandwiching a \ce{LaNiO3} layer between layers of an insulating oxide,
such as \ce{LaAlO3}, will confine the $3d_{3z^2-r^2}$ orbital in the $z$-direction,
removing this band from the FS.
This way, one restricts the electron to the $3d_{x^2-y^2}$ orbital,
similar to the conventional cuprate case.
The DMFT calculation of Reference \cite{hansmann_turning_2009} yields a single-sheet
FS with a small ($30 \%$) $3d_{3z^2-r^2}$ component.
The more recent experimental finding of a \Tc{} of only 3K in related
heterostructures, however, challenges
the hitherto belief that \Tc{} is optimized when the $3d_{x^2-y^2}$ orbital weight
is concentrated in a single band.


Following up on previous work, we compare the high-\Tc{} to the low-\Tc{}
regime of the $e_g$ two-orbital Hubbard model by employing a toolkit
composed of a variety of numerical methods. Our analysis is performed through Kohn-Luttinger (KL) type
calculations~\cite{PhysRevLett.15.524,raghu_superconductivity_2010} in the
weak-coupling and through functional renormalization group (FRG)~\cite{RevModPhys.84.299,platt_functional_2013}
as well as random phase approximation (RPA) \cite{PhysRevLett.17.433,PhysRevB.34.8190,Graser_2009} studies in the
intermediate-coupling regime.
We find that, most importantly, the $e_g$ energy splitting and
the orbital filling
turn out to be crucial parameters to unravel different
superconducting orders, and drastically varying pairing
strengths.
All methods yield $d$-wave and extended $s$-wave pairing as the leading
and subleading superconducting orders in the high-\Tc{} regime. The ordering hierarchy in the low-\Tc{} regime becomes significantly
more susceptible to even a small change of parameters, and hence less universal. 

\textit{\orb{e}{g} minimal model.}---
%
\begin{figure}
	\includegraphics[width=1\columnwidth]{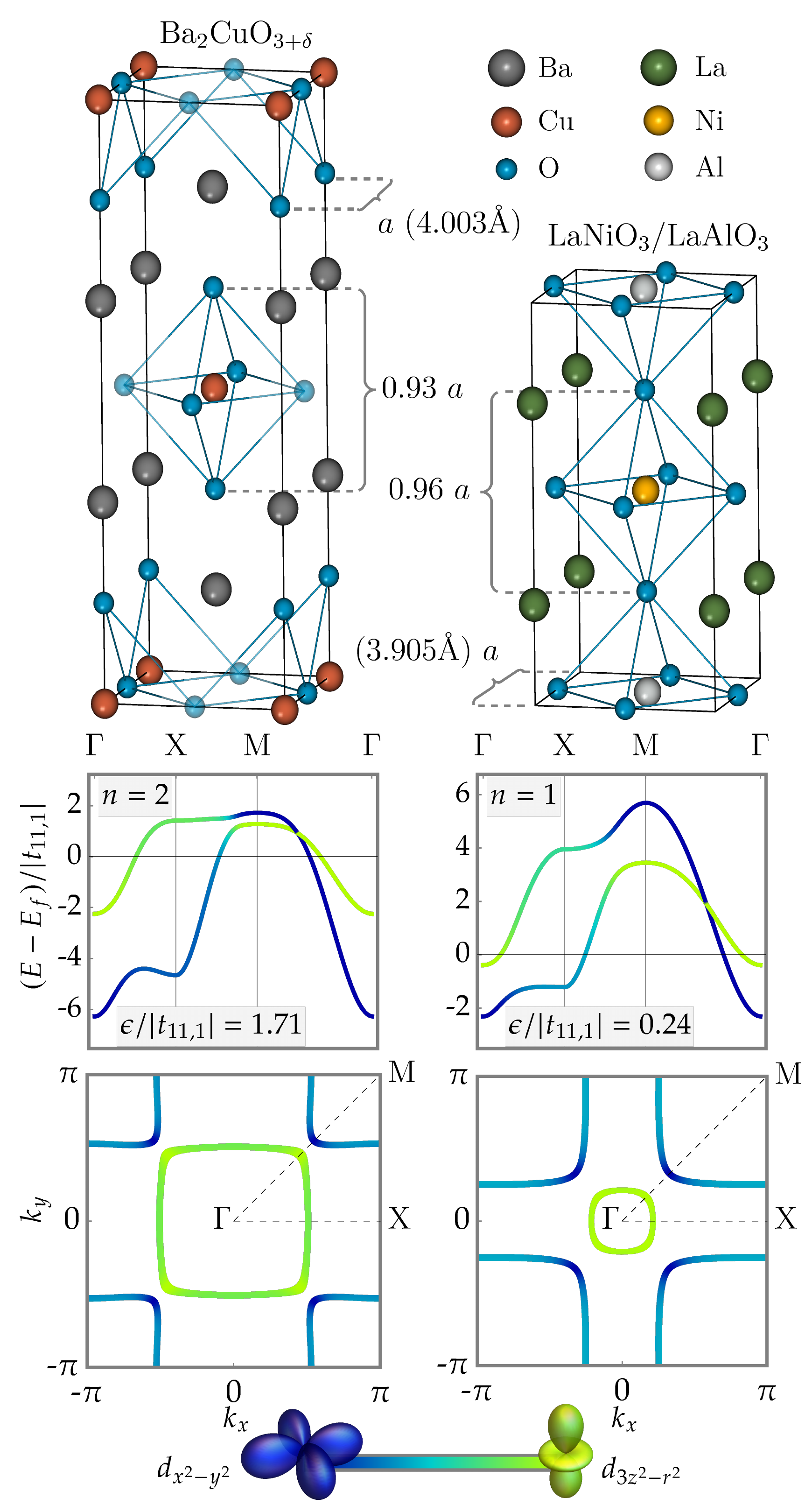}
	\caption{\label{fig:crystal_structure} Sketch of unit cell, band structure and Fermi surface for BCO (unit cell for $\delta=1$, left) and LNO/LAO (right). The orbital weight of a state is given by the respective color blend of blue (\(d_{x^2-y^2}\)) and green (\(d_{3z^2-r^2}\)), we indicate the density of states on the Fermi surface by the thickness of the line.}
	\label{fig:bandstrukturepaper}
\end{figure}
%
%
The shared geometry of the \ce{TO2} (T = Ni, Cu) planes leads to
similarities of the ratios of different transition matrix elements
\(t_{o'o,\alpha}\) between neighbouring sites, as established by ab
initio density functional theory (DFT) calculations (See
  Supplemental Material SM) yielding band structures such as Figure \ref{fig:bandstrukturepaper}.
This allows us to identify two critical parameters that distinguish the high-\Tc{}
material BCO from the low-\Tc{} nickelate heterostructure.
The first is the \(e_g\) manifold splitting \(\epsilon\)
caused by Jahn-Teller distortions of the oxygen octahedra, which is significantly
enhanced in the cuprate system ($0.87$ eV) compared to the nickelate
system ($0.11$ eV).
The second crucial distinction, which is particularly visible through the paradigmatic material limits
cuprates and nickelates, is given by the doping level $\delta$.
While the transition metal ion in both cases is nominally given by \(3d^7\) for
LNO/LAO and \ce{Ba2CuO4}, reducing the oxygen doping in BCO \(\delta\) will
increase the filling of the $d$-shell up to a \(3d^9\) configuration for \ce{Ba2CuO3}.
In our model, this is taken into account by increasing the filling of the \(e_g\) subspace for BCO to \(n=2\)
electrons.
The Hamiltonian we adopt for both materials (albeit with different parameters)
is given by
\begin{align}
	\begin{split}
		H_0 &=
		\sum_{i,\sigma} \sum_{o=1,2}\left( (-1)^o \frac{\epsilon}{2} - \mu \right) c_{o,i,\sigma}^\cre c_{o,i,\sigma}^\ann
		\\&+ \sum_{i,\sigma} \sum_{j\neq i}
		\sum_{o',o=1,2} t_{o'o,\alpha(i,j)} c_{o',i,\sigma}^\cre c_{o,j,\sigma}^\ann \mcomma \label{eq:H0}
		\end{split}
\end{align}
where \(c_{o,i}^\cre\) creates an electron in the orbital
\(o=1,2\) (\(d_{x^2-y^2}\),\(d_{3z^2-r^2}\))
on site \(\vec{r}_i\).
The index \(\alpha(i,j)\) counts the proximity of different sites \(i\) and \(j\)
and is used to label the corresponding orbital hybridizations \(t_{o'o,\alpha}\),
\(\epsilon\) indicates the size of the onsite energy difference for the relevant orbitals
and the chemical potential \(\mu\) controls the filling of the system.
Starting from DFT-calculations \cite{Giannozzi_2009,Giannozzi_2017}
for our prototypical material instances LNO/LAO and BCO,
the tight-binding parameters for this low energy model were obtained by
projecting the DFT result onto a pair of maximally localized Wannier orbitals
on the transition metal ions \cite{Marzari_2012}. 
The resulting model parameters for both LNO heterostructures
and BCO are given in the SM.

Despite the different material chemistry, the similarity of the obtained
parameter sets after normalizing to the bandwidth is remarkable, given
the strongly differing \Tc{}.
We are able to identify two significant differences:
(i)~the orbital splitting and
(ii) the filling fraction of the \orb{$e$}{g} doublet.
While the orbital splitting mainly controls the energy splitting of the bands
at the \(\Gamma\) point as well as the orbital hybridization,
the resulting differences in the Fermiology, seen in Figure \ref{fig:crystal_structure},
result primarily from the difference in the chemical potential.

We model the interaction by adopting a Kanamori type interaction
Hamiltonian,
comprising the four onsite interaction terms in the considered multi-orbital model
\begin{equation}
	\begin{split}
		H_I &= U \sum_{i,o} c_{o,i,\uparrow}^\cre c_{o,i,\uparrow}^\ann c_{o,i,\downarrow}^\cre c_{o,i,\downarrow}^\ann \\&+
		V \sum_{i,\sigma,\sigma'} c_{1,i,\sigma}^\cre c_{1,i,\sigma}^\ann c_{2,i,\sigma'}^\cre c_{2,i,\sigma'}^\ann\\ &+
		J \sum_{i,\sigma,\sigma'} c_{1,i,\sigma}^\cre c_{2,i,\sigma'}^\cre c_{1,i,\sigma'}^\ann  c_{2,i,\sigma}^\ann \\&+
		J' \sum_{i,o\neq o'} c_{o,i,\uparrow}^\cre c_{o,i,\downarrow}^\cre c_{o',i,\downarrow}^\ann c_{o',i,\uparrow}^\ann, \label{eq:HI}
	\end{split}
\end{equation}
with intra-orbital (\(U\)), inter-orbital (\(V\)) repulsive interactions as well as
Hunds coupling (\(J\)) and pair hopping terms (\(J'\)).
The orbital makeup of the BCO bands at the Fermi level are cleaner than the ones of LNO/LAO
due to the larger orbital splitting $\epsilon$. As a consequence,
inter- (intra-) orbital interactions and inter- (intra-) pocket/band
interactions correlate more strongly in the case of BCO compared to
LNO/LAO.

The interaction parameters for the approximation of
a rotationally invariant system are given by $V = U - 2J$ and $J' = J$
\cite{Kanamori_1963,Brandow_1977}.
We further fix $J=0.25~U$ (resulting in $V=U/2$ ) and are subsequently left
with the overall interaction scale \(U\) as the only free parameter.
Since the rotational symmetry of the \(e_g\) complex is broken by the
octahedral crystal field, it is interesting to compare interaction schemes
beyond this simple modelling.
The ratio of inter- versus intra-orbital interaction strength hence is a
reasonable parameter to explore, and we further use it to
gain additional insight into multi-orbital interaction effects in
our effective model.
While actual materials will be limited to the vicinity of \(V/U\approx1/2\),
it is revealing to fully probe the available parameter space, as implemented in the SM.
For the remainder of the main text, we fix $V=U/2$ and choose
the overall interaction scale appropriately
\cite{PhysRevB.84.235121,raghu_superconductivity_2010} (see also SM).


\begin{figure}[t!]
	\includegraphics[width=1.\columnwidth]{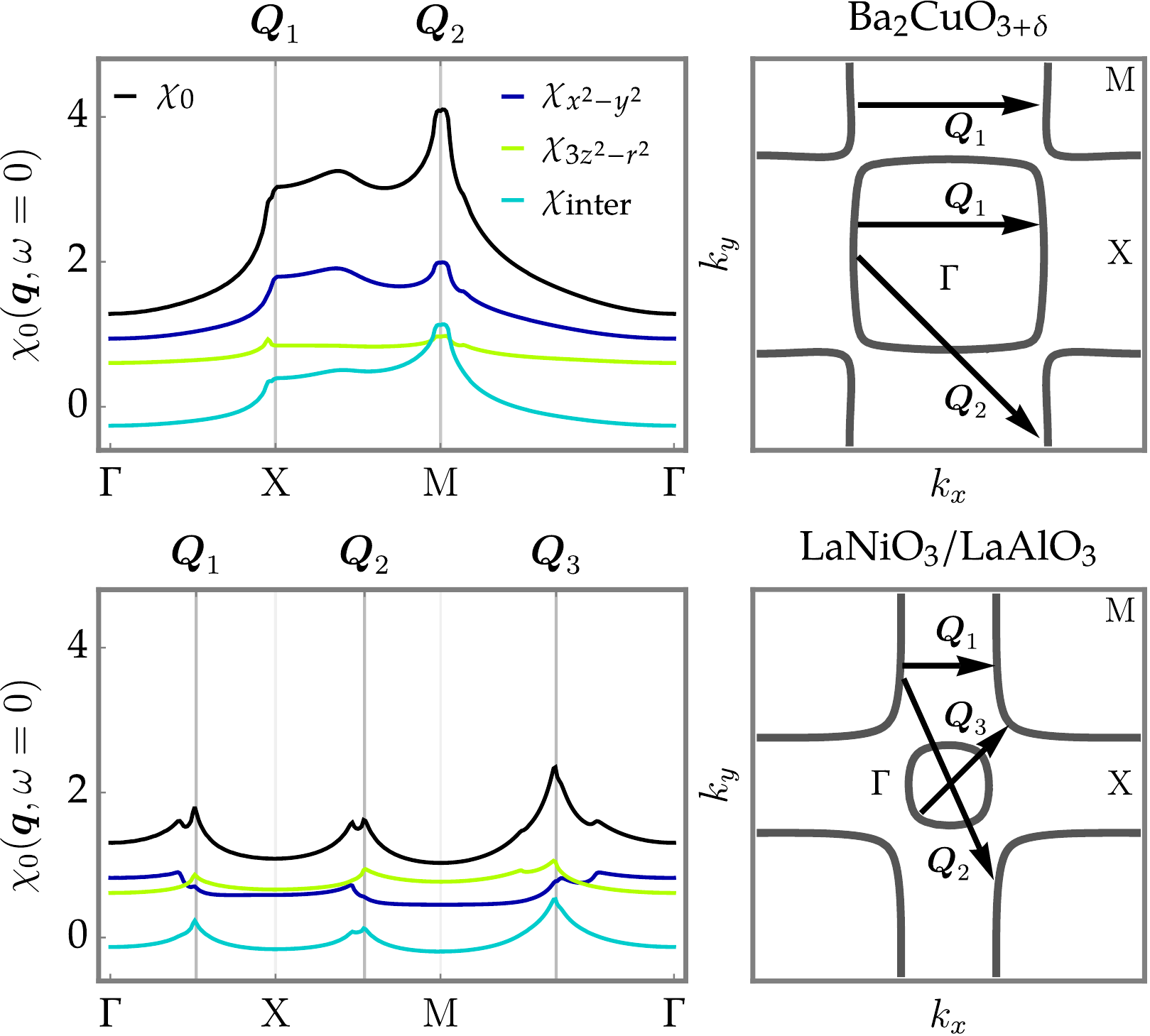}
	\caption{\label{fig:baresusceptibility} Bare susceptibility in arbitrary units for BCO (top panel) and  LNO/LAO (bottom panel) with $\chi^{11}=\chi_{x^2-y^2}$, $\chi^{22}=\chi_{3z^2-r^2}$, and $\chi^{12}+\chi^{21}=\chi_{\mbox{inter}}$ and leading nesting vectors shown in the corresponding BZ (right hand side).}
\end{figure}

\textit{Results.}---In order to get a first impression of the primary spin fluctuation channels, we analyze the bare particle-hole susceptibility
\begin{equation}
	\begin{split}
		\chi_0(\bm{q},\omega)
		&=\sum_{oo'}\sum_{\bm{k},\text{i}\bar \omega_n}\left(G^{oo'}_{\bm{k}+\bm{q},\omega+\text{i}\bar \omega_n}G^{o'o}_{\bm{k},\text{i}\bar \omega_n}\right) \\
		&=\sum_{oo'}\chi^{oo'}(\bm{q},\omega).		
	\end{split}
\end{equation}
It is determined by the single-particle Green's functions $G^{oo'}_{\boldsymbol{k}, \text{i}\omega_n} $ and hence independent of the employed approximations for the treatment of interactions beyond DFT.
We show its zero frequency limit in Figure \ref{fig:baresusceptibility} and highlight corresponding nesting features of the Fermi surface. In contrast to LNO/LAO, which shows an overall uniform bare susceptibility, the nesting in BCO is strongly enhanced for the transfer momenta $\boldsymbol{Q}_1 = (\pi,0), (0,\pi)$ and $\boldsymbol{Q}_2 = (\pi,\pi)$.
These commensurate nesting vectors induce pronounced spin fluctuations in the system which will finally result in attractive interaction channels for the pair-scattering vertex (see e.g. Reference \cite{RevModPhys.84.1383}).

As we analyze the onset of superconductivity for LNO/LAO and BCO in our effective model, this is pursued through different methods:
a Kohn-Luttinger type
analysis~\cite{PhysRevLett.15.524,raghu_superconductivity_2010}, which
is asymptotically exact at infinitesimal coupling, is complemented by
random phase
approximation~\cite{PhysRevLett.17.433,PhysRevB.34.8190,Graser_2009}
as well as functional renormalization
group~\cite{RevModPhys.84.299,platt_functional_2013} calculations that
are usually employed within the intermediate coupling regime.
In the main Letter text, we constrain ourselves mainly to the presentation of the fRG results, and refer the reader to the detailed SM for additional information.
Since the material instances are most likely located in the intermediate coupling regime, the fRG provides the most systematic treatment of superconductivity, as it treats all particle-hole and particle-particle channels on equal footing.
It thus allows to most directly resolve the connection between spin fluctuations channels and superconductivity. Still, one needs to stay aware of the fact that all numerical methods at intermediate coupling are just approximations.
Due to this, we also added the Kohn-Luttinger analysis, in order to have a rigorous reference point at infinitesimal coupling \cite{ShankarWC}.

Within fRG, we determine the critical cutoff energy $\Lambda_c$ and effective two particle irreducible vertex $\Gamma^{\Lambda_c}(\bs k_1,\bs k_2,\bs k_3,\bs k_4)$ for our effective model. As with all other methods used, a short review of the fRG methodology is delegated to the SM.
The renormalization flow breaks down at $\Lambda_c$ \ie the entries of $\Gamma^{\Lambda_c}$ diverge so that we can classify the leading instability by decomposing the full effective vertex $\Gamma^{\Lambda_c}$ into mean-field channels.
For superconductivity, we obtain the effective Cooper pair interaction $\Gamma^{\text{SC},\Lambda_c}(\bs k, \bs k')=\Gamma^{\Lambda_c}(\bs k,-\bs k,\bs k',-\bs k')$.
By solving the linearized gap equation
\begin{equation}
\lambda\Delta_{\bs k}= \sum_{\bs k'}\Gamma^{\text{SC},\Lambda_c}(\bs k, \bs k')\Delta_{\bs k'}
\end{equation}
with \(\Lambda_c \propto T_c\), we identify the symmetry class of the superconducting gap function $\Delta_{\bs k}$ for the smallest eigenvalue $\lambda$ at $\Lambda_c$.

\begin{figure}[t]
	\includegraphics[width=1.\columnwidth]{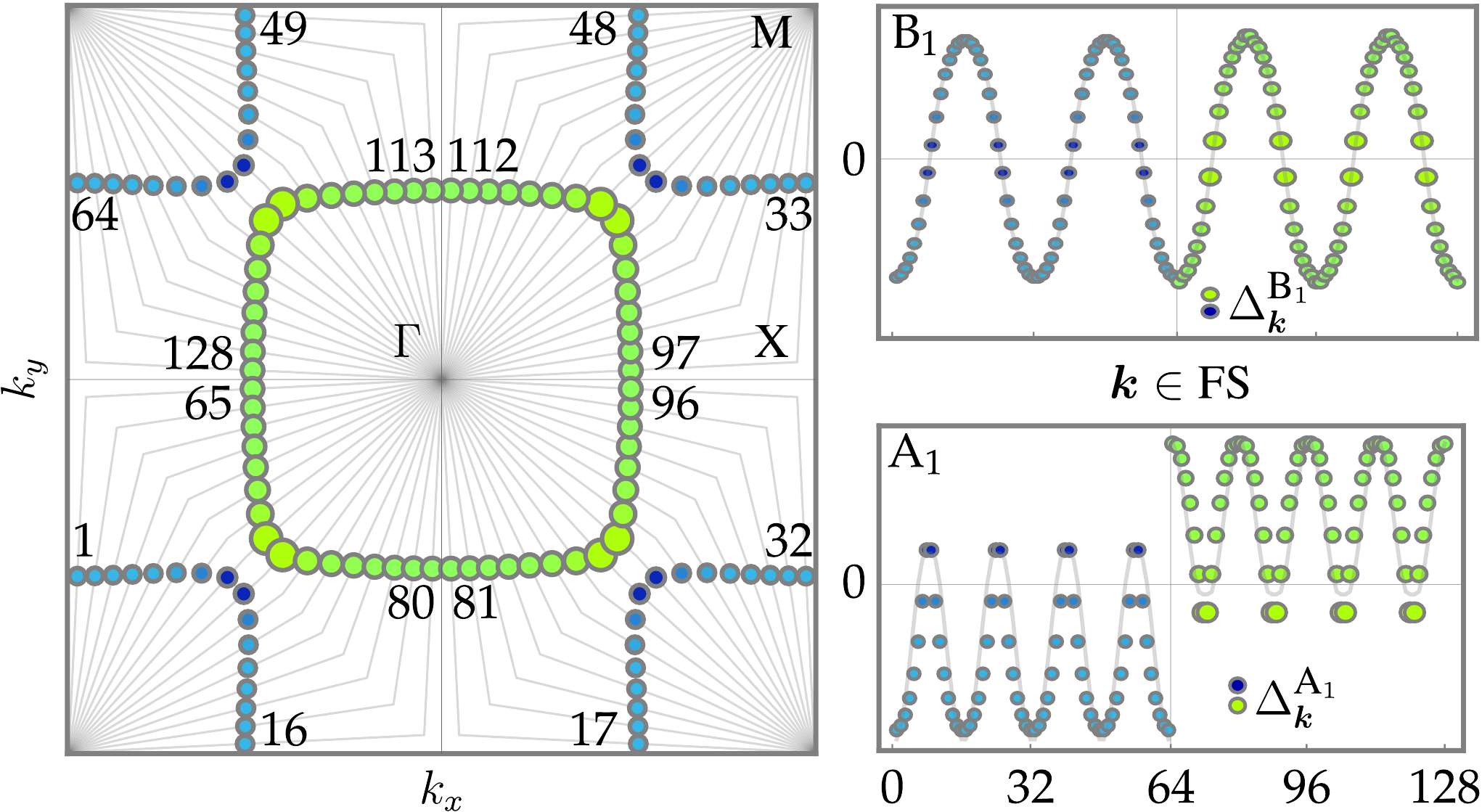}
	\caption{\label{fig:gapplot} Fermi surface (left) of BCO discretized in patches (1-128) for the numerical fRG study. The color scheme, representing the orbital weights, is in accordance with Figure \ref{fig:crystal_structure} and the thickness of the plot marker indicates the DOS. On the right hand side we plotted the form factor of the leading (upper panel) and sub leading (lower panel) eigenvalue $\lambda$ for the superconducting channel on the discretized Fermi surface at the break down of the fRG flow. They transform according to the $B_1$ and $A_1$ irreducible representations of the lattices \(C_{4v}\) point group symmetry respectively. The fitted harmonic fingerprint of these form factors is indicated by the grey lines in the panels on the right.}
\end{figure}

Figure \ref{fig:gapplot} displays the gap functions for the leading
and subleading eigenvalues $\lambda$ for BCO. We find a cuprate-like
$d_{x^2-y^2}$-wave form factor for the leading eigenvalue of
$\Gamma^{\text{SC},\Lambda_c}(\bs k,\bs k')$ and extended $s$-wave order for the sub-leading eigenvalue. Assigning the associated irreducible lattice representation (irrep), the leading eigenvalue possesses the symmetry character $B_1$ with $\Delta_{\bs k}^{B_1} \propto (\cos k_x- \cos k_y) +0.001~(\cos 2k_x- \cos 2k_y)$ and the subleading one $A_1$ with
$\Delta_{\bs k}^{A_1} \propto (\cos k_x + \cos k_y) + 0.45~(\cos 3k_x + \cos 3k_y)  $. The harmonic decomposition of these form factors is obtained by fitting to the fRG eigenvectors. For BCO, these results agree with the RPA results presented in the SM~\cite{Note1}, which in turn reproduce previous RPA results by Maier \ea~\cite{maier_d_wave_2018}.

It is quite transparent in fRG, how these superconducting orders relate to the pronounced spin fluctuation channels: the $B_1$ and $A_1$ form factors are enhanced by the dominant pair scattering process with momentum transfer $\boldsymbol Q_2$, and partially by $\boldsymbol Q_1$. 
The corners of the Fermi surface are favourably nested to a majority of the momenta on the same Fermi pocket via $\boldsymbol Q_1$ and the corners of the other Fermi pocket via $\boldsymbol Q_2$, yielding additional nodes and a particularly anisotropic $A_1$ form factor.

\begin{table}[t!]
	\begin{tabular}{
			l
			c 
			c 
			c 
		}
		\toprule
\multirow{2}{*}{Method~} &  \multirow{2}{*}{\( \text{Log}_{10}\left(\ddfrac{\text{\Tc}^{\ce{LNO/LAO}}}{\text{\Tc}^{\ce{BCO}}} \right)\)} & \multicolumn{2}{c}{(sub-)leading  irrep} \\
		&                                                                 &   {\ce{BCO}  \( \quad\)}   & {\ce{LNO/LAO}}  \\
		\midrule
		fRG & {\(\mathcal{O}(-2)\)} 
			   & {\( B_1 \) (\( A_1 \))  \( \quad\)} 
			   & {\( E_{\phantom{2}} \) (\( B_2 \))} \\
		RPA & {\(\mathcal{O}(-10)\)} 
		 	   & {\( B_1 \) (\( A_1 \))   \( \quad\)} 
		 	   & {\( B_2 \) (\( B_1 \))} \\
		KL & {\(-408(\text{eV}^2)/U^2\)}
		     & {\( A_1 \) (\( B_1 \))  \( \quad\)}
		     & {\( E_{\phantom{2}} \) (\( B_2 \))} \\
		\bottomrule
	\end{tabular}
	\caption{
		Ratio of the critical temperature for LNO/LAO and BCO calculated by fRG (functional RG), RPA (Random Phase Approximation), and KL (Kohn-Luttinger) analysis and the corresponding classification of the gap function in irreducible lattice representations (irreps). The irreps are often named according to their nodal structure: \(A_1\) is referred to as an (extended) \(s\)-wave, \(E\) as a \(p\)-wave and \(B_1\)(\(B_2\)) as \(d_{x^2-y^2}\)(\(d_{xy}\)).}
	\label{tab:ratioTc}
\end{table}

By contrast, the fRG analysis of LNO/LAO yields an upper bound for \Tc{} which is two orders lower in
magnitude, and a form factor transforming under the $E$ irrep of the
crystal's point group. The Fermi pockets feature no particularly pronounced nesting which,
combined with the unclean orbital makeup, yields a nearly uniform pairwise interaction
between different points of the Fermi surface. As a consequence, $\Lambda_c$ is dramatically decreased compared to the BCO results.

We present the symmetry class of the leading form factor and the ratio
of \Tc{} for BCO and LNO/LAO in Table \ref{tab:ratioTc} for the fRG,
RPA, and KL calculations. A unanimous finding of all methods is the overall trend of lower \Tc{} for LNO/LAO as well as the leading $d_{x^2-y^2}$- and $s$-wave instabilities for the high-\Tc{} case in BCO.
For LNO/LAO, all methods substantially differ from each other. Given the small instability scale and, from there, the enhanced sensitivity of the result to the specific formulation of the approximative method, this is not surprising.
It shows, that in the low-\Tc, or rather pairing noise, regime, even slight biases of different approximation schemes manage to affect the eventual result, and strongly enhances the volatility of any result for the superconducting instability. Nevertheless, it is still interesting to trace back the biases of the different methods in such a case, which is delegated to the SM.


\textit{Conclusion.}---Already in an effective description as simple as the $e_g$ minimal model studied in this work, we can identify the enormous range of multi-orbital effects on \Tc{} at the example of a multi-orbital high-\Tc{} material \ce{Ba2CuO_{3+\delta}} and a multi-orbital low-\Tc{} material LaNiO$_3$/LaAlO$_3$. Certainly, this study is not exhaustive in describing all multi-orbital effects in transition metal superconductors. For instance, not only multiple orbitals of the transition metal atom, but also other orbital degrees of freedom may prevail at low energies, such as recently observed for infinite layer nickelates. Still, we expect the minimal modelling of multi-orbital effects to constitute a promising future direction to close the gap between experimental evidence and theoretical simulation of unconventional superconductors.


\textit{Acknowledgments.}---This work is funded by the Deutsche Forschungsgemeinschaft (DFG, German Research Foundation) through Project-ID 258499086 - SFB 1170 and through the W\"{u}rzburg-Dresden Cluster of Excellence on Complexity and Topology in Quantum Matter-ct.qmat Project-ID 390858490 - EXC 2147.
SR acknowledges support from the Australian Research Council through FT180100211 and DP200101118. 
We further gratefully acknowledge the Gauss Centre for Supercomputing e.V. \cite{GaussCenter} for providing computing time on the GCS Supercomputer SuperMUC at Leibniz Supercomputing Centre \cite{SuperMUC} and the HPC facility Spartan hosted at the University of Melbourne.

\clearpage

\onecolumngrid
\vspace{\columnsep}
\begin{center}
\large
 \textbf{Supplemental Material\\
	From high-$T_c$ to low-$T_c$: Multi-orbital effects in
	transition metal oxides}

\end{center}
\vspace{\columnsep}
\twocolumngrid

\par

\begin{figure*}[t!]
	\centering
	\includegraphics[width=1.85\columnwidth]{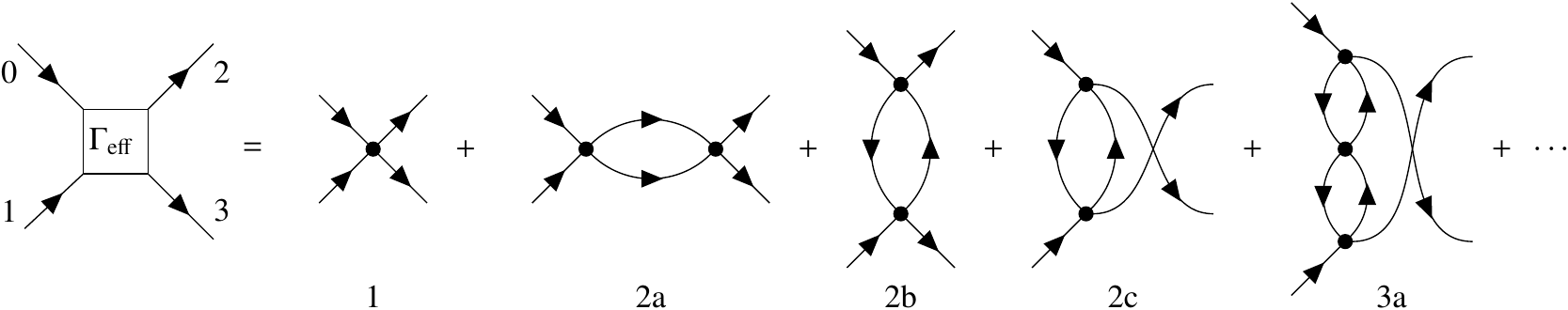}
	\caption{
		\label{Fig:App:diagram}Diagrammatic expansion of the effective two-particle-vertex $\Gamma_{\text{eff}}$ in perturbation theory. Internal lines represent fermionic propagators and solid points represent the bare two particle interaction $\Gamma_{b}$. There are three topological different diagrams up two second order in $\Gamma_{b}$. We absorb all additional factors and signs in their respective diagram.
		Depending on the method different diagrams are chosen as explained in the corresponding subsections.}
	\label{fig:diagramsexpansion}
\end{figure*}


\section{First-principles analysis}
We performed first-principles
calculations within the framework of the density functional
theory (DFT) as implemented in the in the
QUANTUM ESPRESSO suite~\cite{Giannozzi_2009,Giannozzi_2017}.
The generalized gradient approximation, as parametrized by the Perdew-Burke-Ernzerhof
generalized gradient approximation (PBE-GGA) functional for the exchange-correlation potential, was used by expanding
the Kohn-Sham wave functions into plane waves up to an
energy cutoff of \(100\,\text{Ry}\approx1.4\,\)keV
and sampling the Brillouin zone on a $4 \times 4 \times 2$ regular mesh~\cite{PhysRevLett.77.3865}.
For \ce{Ba2CuO4} we used the crystal structure parameters available from experiment~\cite{Li12156} (\(a=4.003\,\)\AA, \(c=12.94\)\AA).
Due to the absence of experimental data on the the \ce{LaAlO3}/\ce{LaNiO3} heterostructure (LAO/LNO) proposed in \cite{chaloupka_orbital_2008,hansmann_turning_2009},
we simulated its epitaxial growth by imposing an in-plane lattice constant of \(a=3.905\,\)\AA~and relaxing the out of plane lattice parameter as well as the
atomic positions on a refined $24 \times 24 \times 24$ regular mesh Brillouin zone mesh, resulting in \(c=7.500\,\)\AA.
The extraction of the two-orbital minimal
model used in the presented analysis of superconducting instabilities in
both materials was based on the Wannier functions formalism \cite{Marzari_2012}
and we fixed the Fermi level in our model by enforcing the filling in the
simplified model as opposed to the DFT Fermi level.

\section{Minimal model}
The minimal model to describe the low energy physics of both materials studied here is an extended two-band Hubbard model
on a square lattice. While this becomes apparent from the performed DFT calculations,
one can also arrive at this conclusion by simple chemistry considerations:
Both the LAO/LNO and \ce{Ba2CuO_{3+\delta}} feature a layered perovskite structure.
In case of LNO/LAO nickel oxide planes \ce{Ni^{3+}O2^{2-}} alternate
with an insulating aluminium oxide plane \ce{Al^{3+}O2^{2-}}, allowing
for a two-dimensional description of the low energy electronic degrees
of freedom located on the nickel ions
\cite{hansmann_turning_2009}. For BCO, copper oxide planes
\ce{Cu^{(2+2\delta)+} O2^{2-}}, which are pairwise shifted against
each other, alternate with two insulating barium oxide planes
\ce{Ba_{2}^{2+}O^{2-}}. As such, both systems yield a two-dimensional
description of the low energy electronic degrees of freedom located on
the copper or nickel ions \cite{maier_d_wave_2018,Li12156}.

As it is typical for perovskite structures, the transition metal ion is enclosed by an octahedron of
\ce{O^2-} ions. This results in a crystal field which splits the \(L=2\) d-orbital quintuplet
into a low energy \(t_{2g}\) triplet (composed of the \(d_{xy}\), \(d_{xz}\) and \(d_{yz}\) orbitals)
and an \(e_g\) doublet (composed of the remaining \(d_{x^2-y^2}\) and \(d_{3z^2-r^2}\)) with higher energy.
For both LNO/LAO (\ce{La^{3+}Ni^{3+}O3^{6-}}/\ce{La^{3+}Al^{3+}O3^{6-}}) and \ce{Ba2CuO_{4}} (\ce{Ba2^{4+}Cu^{4+}O4^{8-}})
the transition metal ions nickel and copper have a \(3d^7\) electronic configuration, while the \(3d^6\) configuration
on the aluminum site of LNO/LAO results in a significant excitation gap allowing us to remove it from the low energy description of the system.
In the case of BCO, we additionally simulate the oxygen deficiencies by artifically raising the Fermi level in our model compared to the \ce{Ba2CuO4} calculation, simulating the effect
of the added electrons by removing oxygen from the compound.

The completely filled \(t_{2g}\) multiplett is split of by the octahedral crystal field,
allowing us to restrict our analysis to the \(e_g\) doublet.
Additionally, degeneracy of the \(e_g\) states is lifted by distortions of
the oxygen octahedron. The strength of this Jahn-Teller distortion and the
resulting \(e_g\) splitting \(\epsilon\) is one of the key parameters resolved
in our work. We find our effective low-energy model to be formed by a two band Hubbard model
on a square lattice spanned by the transition metal's \(d_{x^2-y^2}\) (\(o=1\)) and \(d_{3z^2-r^2}\) (\(o=2\)) orbitals:
\begin{align}
\begin{split}
H_0 &=
\sum_i \sum_{o=1}^2 \left( (-1)^o \frac{\epsilon}{2} - \mu \right) c_{o,i}^\cre c_{o,i}^\ann
\\&+ \sum_{i} \sum_{j\neq i}
\sum_{o',o=1}^2 t_{o'o,\alpha(i,j)} c_{o',i}^\cre c_{o,j}^\ann \label{eq:appH0}
\end{split}
\\&=  \sum_{i} \sum_{k}
\sum_{o',o=1}^2 \xi_{o'o}(\vec{k}) c_{o',k}^\cre c_{o,k}^\ann \label{eq:appH0k}
\end{align}
with the single-particle bands
\begin{align}
\label{eq:bloch_intra}
\xi_{oo}(\vec{k})&=\left( (-1)^o \frac{\epsilon}{2} - \mu \right) +2t_{oo,1}(\cos k_{x}+\cos k_{y})\nonumber\\
&+4t_{oo,2}\cos k_{x}\cos k_{y}+2t_{oo,3}(\cos 2k_{x}+\cos 2k_{y})\nonumber,\\[10pt]
\xi_{oo'}^{o\neq o'}(\vec{k})&=\,2t_{oo',1}(\cos k_{x}-\cos k_{y})\nonumber\\[5pt]
&+2t_{oo',3}(\cos 2k_{x}-\cos 2k_{y})\nonumber,
\end{align}
where \(c_{o,i}^\cre\) creates an electron in the orbital \(o=1,2\) ($d_{x^2-y^2}$, $d_{3z^2-r^2}$) on site
\(\vec{r}_i\).
The index \(\alpha(i,j)\) counts the proximity of different sites \(i\) and \(j\)
and is used to label the corresponding orbital hybridizations \(t_{oo,\alpha}\).
The chemical potential \(\mu\) controls the filling of the system and \(\epsilon\) is the relative onsite energy of the relevant orbitals.
Tab.\,\ref{tab:parameters} lists the tight binding parameters for the minimal model that we obtained by fitting this model to the
results from our DFT calculations.

\begin{table}[!h]
	\begin{tabular}{
			l
			S[table-figures-integer = 1, table-figures-decimal = 4, table-number-alignment = center]
			S[table-figures-integer = 1, table-figures-decimal = 4, table-number-alignment = center]
			S[table-figures-integer = 1, table-figures-decimal = 4, table-number-alignment = center]
		}
		\toprule
		%

		model parameters           & {\ce{Ba2CuO_{3+\delta}}}   
		& {\ce{LaNiO3/LaAlO3}}   \\
		\midrule
		{\(t_0\, / \)\,\si{\eV}}   & -0.511                     
		& -0.447                 \\
		{\(n\, / \)\,electrons}    &  2.00                      
		&  1.00                  \\
		\midrule
		{\(\epsilon\, / \,|t_0|\)}   & 1.71                      
		& 0.24                  \\
		{\(\mu\, / \,|t_0|\)}        & +0.82                     
		& -1.81                  \\
		\midrule
		\multicolumn{2}{l} {hopping amplitudes \(/\,|t_0| \) }    &                            &                        \\
		\midrule
		{\(t_{11,\alpha=1}\)}      & -1.00                      
		& -1.00                  \\
		{\(t_{22,\alpha=1}\)}      & -0.44                      
		& -0.48                  \\
		{\(t_{12,\alpha=1}\)}      & +0.63                      
		& +0.62                  \\
		\midrule
		{\(t_{11,\alpha=2}\)}      & +0.13                      
		& +0.13                  \\
		{\(t_{22,\alpha=2}\)}      & -0.07                      
		& -0.07                  \\
		{\(t_{12,\alpha=2}\)}      & +0.00                      
		& +0.00                  \\
		\midrule
		{\(t_{11,\alpha=3}\)}      & -0.28                     
		& -0.13                  \\
		{\(t_{22,\alpha=3}\)}      & -0.06                      
		& -0.03                  \\
		{\(t_{12,\alpha=3}\)}      & +0.11                      
		& +0.06                  \\
		\bottomrule
	\end{tabular}
	\caption{
		Parameters for the tight binding Hamiltonian \eqref{eq:appH0} for both LNO/LAO and BCO. \(t_{oo',\alpha}\) denotes the hybridization between two orbitals \(o\) and \(o'\) that are
		\(\alpha\)\textsuperscript{th} nearest neighbours (\(\alpha = 1\) corresponds to nearest neigbour, \(\alpha = 2\) to next-nearest neigbour etc.).
	}
	\label{tab:parameters}
\end{table}
%


\section{Methodology}
In this section we aim to give a short overview on different diagrammatic schemes to calculate the effective two-particle vertex function, the central object to reveal the superconducting instabilities.
Specifically, we perform Kohn-Luttinger type calculations\,\cite{PhysRevLett.15.524,raghu_superconductivity_2010} in the weak coupling (referred to as weak-coupling renormalization group Kohn-Luttinger analysis (KL) in the following) and through functional renormalization group (FRG)\,\cite{RevModPhys.84.299,platt_functional_2013} as well as random phase approximation (RPA)\,\cite{Altmeyer2016} in the intermediate coupling regime.
Fig.\,\ref{Fig:App:diagram} shows the perturbative expansion of the effective two-particle vertex $\Gamma_{\text{eff}}$ in terms of the bare two-particle vertex $\Gamma_b$. We will show how one can use these diagrams to calculate $\Gamma_{\text{eff}}$ within KL analysis and RPA. Further we give a short derivation of the fRG method, which features the same topological diagrams as the KL analysis, in order to solve a differential equation to determine $\Gamma^{\Lambda_c}$, where $\Lambda_c$ is the critical energy scale of the system.
Clearly all methods employed here include different sets of diagrams, resulting in a significant difference of captured screening effects. Therefore it is necessary to taylor the initial interaction strength $U$ for each approximation scheme individually, in order to obtain comparable results.
A more detailed Discussion is presented in the References \cite{PhysRevB.84.235121,raghu_superconductivity_2010}.
Here we choose \(U = 3.5\,\mbox{eV}\) for the fRG calculations and \(U = 0.8\,\mbox{eV}\) for the RPA calculation and therefore account for the additional screening processes include in the fRG approximation.


\subsection{Functional Renormalization Group}
\textit{Method.}---The here presented summary of the FRG method follows Ref.\,\onlinecite{platt_functional_2013}, which provides a detailed derivation of the flow equation.
First we set up our theory at a high energy scale $\Lambda_{\text{init}}$ with the action
\begin{equation*}
S^{\Lambda_{\text{init}}} = S_0 +S_I ,
\end{equation*}
containing the bare action $S_0$ and an interaction term $S_I$.
In order to connect to lower energy scales we integrate out high-energy modes step by step, adjusting the action accordingly and generating a trajectory between microscopic theory at high energies and an effective low-energy description.
We achieve this by introducing a flow parameter $\Lambda$ interpolating the action from the trivial stating point ($\Lambda = \Lambda_{\text{init}}$) to the fully interacting theory ($\Lambda = 0$).
One can deduce an set of integro-differential equation for all irreducible $2n$-point vertex functions. It turns out that the hierarchy of these so called flow equations does not close, meaning that in order to calculate the flow for an $2n$-point vertex function, one has to know the
vertex function of the next higher order. Consequentially, to numerically solve the integro-differential equation, the flow equation for the four-point vertex function is truncated by neglecting all terms containing higher order vertex functions \cite{RevModPhys.84.299,Salmhofer_fRG}. Also we neglect any corrections
of the self energy $\Sigma$ under the flow. The remaining integro-differential equation
is of first order. The derivative of the irreducible four-point vertex function $\Gamma^{(4),\Lambda}_{k_1',k_2';k_1,k_2}$ with respect to $\Lambda$ turns out to be equal to diagrams which are topologically equivalent to $2a-2c$ in Fig.\,\ref{Fig:App:diagram}, the only difference being that one of the internal Green's functions $G^\Lambda$ gets promoted to a
single scale propagator $S^\Lambda = \partial_\Lambda G^\Lambda|_{\Sigma=\text{const}}$.
\begin{widetext}
	\begin{align*}
	\frac{d}{d\Lambda}\Gamma^{(4),\Lambda}_{k_1',k_2';k_1,k_2}   = \sum_{\substack{k,k' \\ q,q'}}G^\Lambda_{k,k'}S^\Lambda_{q,q'}\Big ( \Gamma^{(4),\Lambda}_{k_1',k_2';k,q} \Gamma^{(4),\Lambda}_{k',q';k_1,k_2}
	&- \big [\Gamma^{(4),\Lambda}_{k_1',q';k_1,k} \Gamma^{(4),\Lambda}_{k',k_2';kq,k_2} + (k \leftrightarrow q, k' \leftrightarrow q')\big]  \\
	&+ \big [\Gamma^{(4),\Lambda}_{k_2',q';k_1,k'} \Gamma^{(4),\Lambda}_{k',k_1';q,k_2} + (k \leftrightarrow q, k' \leftrightarrow q')\big] \Big).
	\end{align*}
\end{widetext}
An index $k$ contains the spin $\sigma$, orbital (band) $o$ ($b$), momentum $\boldsymbol k$ and frequency $k_0$ degrees of freedom; the $k$ index on the left corresponds to an in-going particle, while the one on the right to an out-going particle. We further simplify the numerical effort by projecting all momenta to the Fermi surface and compute the flow neglecting all finite frequency contributions.

There are multiple schemes to implement the flow parameter. Wilson's original idea of integrating out momentum modes shell by shell offers one possible implementation \cite{PhysRevA.8.401,POLCHINSKI1984269,WETTERICH199390}, by introducing an energy cutoff depending on $\Lambda$ in the propagators. However this cutoff scheme is
not suited to treat particle-hole fluctuations in an unbiased way\,\cite{platt_functional_2013}. Alternatively, temperature can be used as the flow parameter, as utilized in this work. Doing so not only avoids any issues with the particle hole fluctuations but also offers a very intuitive picture of understanding
the flow as a whole. Integrating out energy modes, descending from high to low energies, can now be interpreted as cooling down our system.

Using the Euler method, we can solve the integro-differential equation. The initial value for the action is chosen to be equivalent to the bare action
\begin{equation}
S^{\Lambda_{\text{init}}} = S_0
\end{equation}
and the initial four-point vertex function at $\Lambda_{\text{init}}$ is set to be the interaction term of the unrenormalized theory, which than flows under the
set up formalism either to an new effective function at $\Lambda=0$ or to a fix point at a given $\Lambda_c$ at which the vertex diverges and the flow breaks down. The later case marks the breakdown of the Fermi surface to a new symmetry broken phase. The order parameter of
the new phase can be analyzed by decomposing the full vertex function into mean field channels. The leading instability is then given by the most diverging of these channels. Analogously to RPA and KL analysis one can further characterize the order parameter according to its transformation behavior under the lattices symmetries, as demonstrated in the main paper.

%

\begin{figure}[t!]
	\centering
	\includegraphics[width=1.\columnwidth]{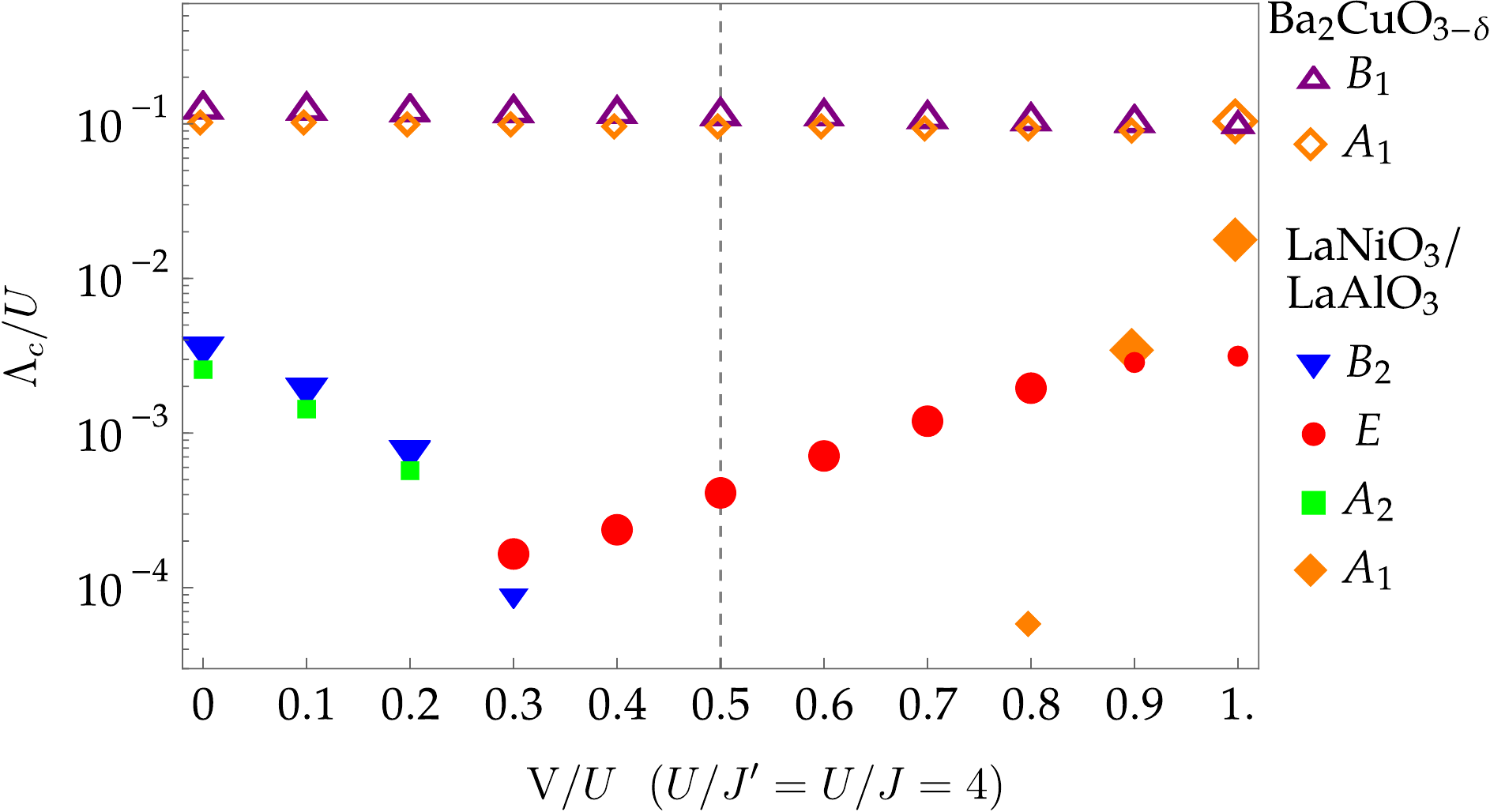}
	\caption{\label{fig:Phasediagramm} Critical cutoff $\Lambda_c$ of fRG-flow and symmetry character of leading and subleading superconducting order parameter for for BCO (unfilled markers) and LNO/LAO (solid markers). The cutoff for the subleading  order parameter is estimated by extrapolating the divergence in $\Lambda$ beyond $\Lambda_c$ of the corresponding eigenvalue $\lambda$ . The interaction is given by $U=4J=4J'=3.5\,\mbox{eV}$ while the different values for $V/U$ is achieved by increasing \(V\). The dashed line indicates the  generic interaction ratio $U=2V=4J=4J'$.}
\end{figure}

\textit{Results.}---Fig.\,\ref{fig:Phasediagramm} shows the critical cutoff energy $\Lambda_c$ for the two relevant models.
The superconducting channel of the meanfield decoupled effective vertex $\Gamma^{Sc,\Lambda_c}(k,k')$ at $\Lambda_c$ can further be analysed by solving the eigenvalue equation
\begin{equation}
\Delta_k\lambda= \sum_{k'}\Gamma^{Sc,\Lambda_c}(k,k')\Delta_{k'} \quad \text{with} \quad \Lambda_c \propto T_c
\end{equation}
and identifying the symmetry class of the from factor $\Delta_k$ for the smallest eigenvalue $\lambda$.

By doing so we can identify two regimes: The first being the case of $V \geq U$, a regime featuring strong repulsive interaction between the pockets of the Fermi surface (FS), which are dominated by opposite orbital character, resulting in an extended $s$-wave ($A_1$ irrep), featuring nodes between the pockets.
The other being $V\ll U$, where intra-pocket scattering makes up for nearly all scattering processes and thus features nodal lines intersecting the FS. In case of \ce{Ba2CuO_{3+\delta}} the leading symmetry class is given by $B_1$ irrep ($d_{x^2-y^2}$-wave), which is the well known  $d_{x^2-y^2}$-wave present in most cuprates, to accommodate the large intra orbital nesting $q=(\pi,\pi)$. For \ce{LaNiO3 / LaAlO3}  the nature of the incommensurate nesting vectors
would favour nodes between the parallel lines of the outer and inner pocket of the Brillouin zone (BZ). The resulting nodal configuration can be accommodated by a form factor of the symmetry class $B_2$ ($d_{xy}$-wave) and the subleading form factor of the symmetry class $A_2$ ($g$-wave).

In between these two regimes the critical cutoff for \ce{Ba2CuO_{3+\delta}} is nearly constant. By increasing the inter orbital interaction nesting between the parallel lines of the FS of different pockets, connected by transition vectors around $\bs q =(\pi,\pi)$, become relevant, hence the symmetry remains $B_1$. Note that the nesting vectors around  $\bs q =(\pi,\pi)$ are yielding favourable results in the gap equation for the $B_1$ and $A_1$ order parameters.
Hence, the transition between these two phases by increasing inter-orbital interactions is quite generic: one just chooses between the nodal configuration which achieves a sign change between the maximal subset of nesting vectors weighted with their interaction strength. This can also be seen in the weak coupling analysis below and in the RPA results from Maier {\it et al.}\,\cite{maier_d_wave_2018}, where the eigenvalues associated with $B_1$ and $A_1$ irreps are nearly degenerate. In fRG these two irreps make up the leading and first subleading symmetry classification of the order parameter for all probed values of $V/U$ with a flow that indicates nearly simultaneous divergences of the respective eigenvalues.

In contrast, for \ce{LaNiO3/LaAlO3}  there is no such sequence of phases sharing advantageous nesting options at the vicinity of the phase transition. Its intermediate interaction regime offers no clear favoured nesting option and its unclean orbital makeup yields a uniform interaction between different points of the FS. Hence, it features a dramatically decreased $\Lambda_c$ and a form factor of the symmetry class $E$ ($p$-wave). This decrease is remarkable for two reasons. Firstly, it is counter intuitive in the sense that we add interaction to the system and decrease \Tc{} by up to two orders of magnitude. Secondly, the resulting form factor of symmetry class $E$ has two representations (\ie the irrep is two-dimensional), which can be added in such a way that the resulting order parameter leads to a hard gap. The $B_2$ $d$-wave is demoted to subleading irrep in this regime, due to the large additional interaction between the two Fermi pockets, which is not aligned with a nodal configuration provided by a $d_{xy}$ order parameter. For $V\approx 0.7~U$ the $s$-wave becomes the first sublading irrep.
%
\begin{figure}[t!]
	\centering
	\includegraphics[width=1.\columnwidth]{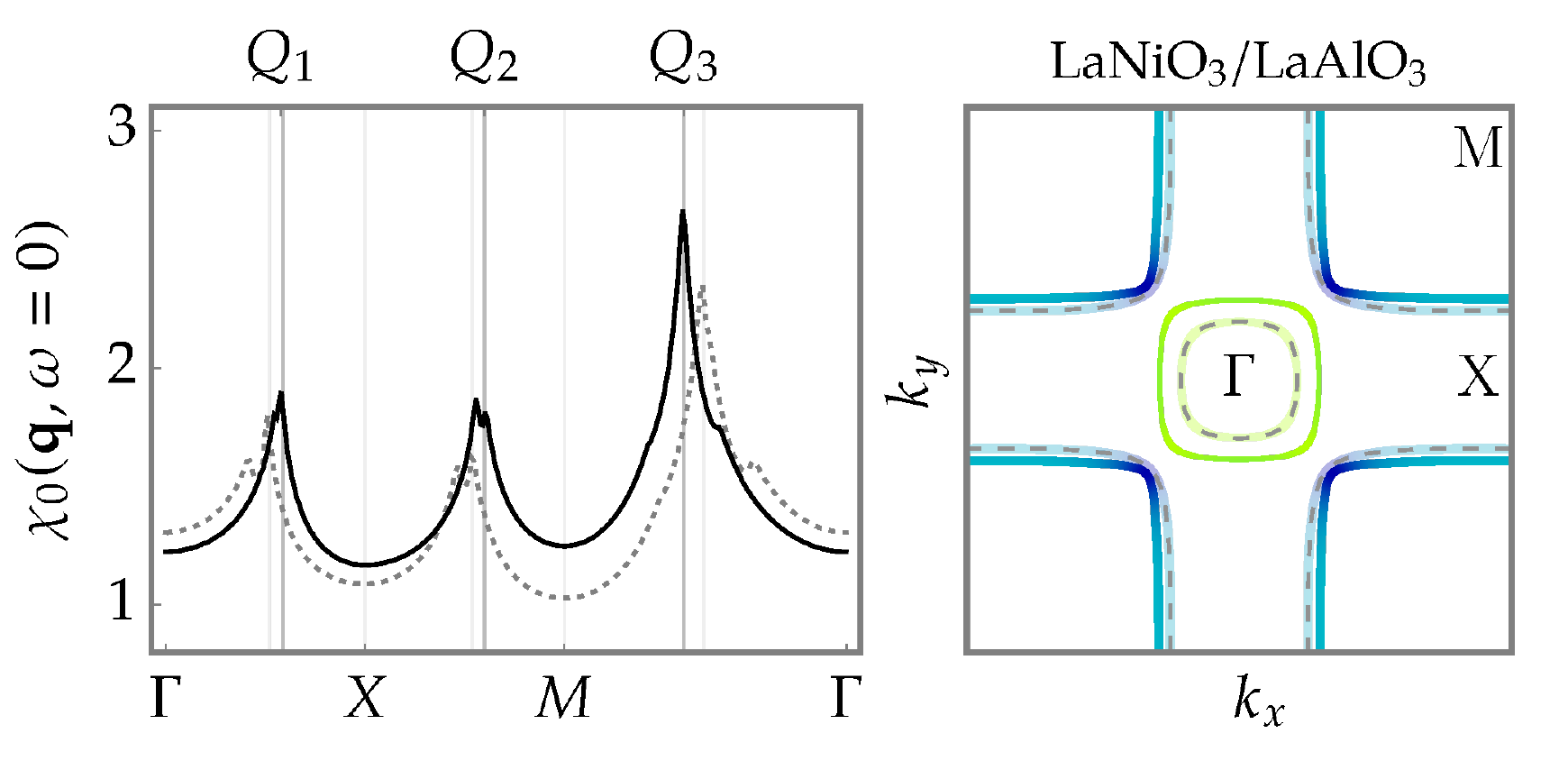}
	\caption{\label{fig:Comparison_FS} Bare susceptibility and corresponding FS for \ce{LaNiO3/LaAlO3}  at filling $n=1.1$ (solid lines) and $n=1.0$ (dashed line). The respective critical cutoffs for the generic interaction ratio $U=2V=4J=4J'$ and $U=3.5\,\mbox{eV}$ are given by $\Lambda_c/U = 7.1\cdot 10^{-3}$ ($n=1.1$) and $\Lambda_c/U = 4.2\cdot 10^{-3}$ ($n=1.0$).}
\end{figure}
%
Overall it seems very reminiscent to frustration effects in magnetism, although this is not related to the geometry of the system, but instead to the relative mixture of different interactions.  Following that line of reasoning one can think of \ce{LaNiO3 / LaAlO3} as a {\it frustrated superconductor}.

Electronic structure calculations performed in Ref. \cite{hansmann_turning_2009}, suggested a larger inner FS pocket compared to our results. Increasing the filling of the presented minimal model we find a similar FS (compare Figure \ref{fig:Comparison_FS}). The critical cutoffs fRG $\Lambda_c$ of both fillings are of the same order of magnitude and yield a form factor of symmetry class $E$, thereby demonstrating the independence of the presented many body results from this band structure detail.

\subsection{Random Phase Approximation}

\textit{Method.}---The RPA formulation is based on the idea that selected particle-hole scattering events add up coherently, whereas all other possible scattering channels are suppressed by acquiring random relative phases. Neglecting these terms in the calculation of the effective two-particle vertex results in a summation up to infinite order of pure bubble and ladder diagrams. For the single orbital case both interactions and susceptibilities are scalar. In contrast the multiorbital case the pairing vertex has an additional contraction over the orbital degrees of freedom. This orbital makeup induces additional diagrams with the structure of vertex corrections which are included in the matrix-RPA formulation\,\cite{Altmeyer2016}. A graphic representation of the considered terms are the particle-hole diagrams shown in Fig.\,\ref{Fig:App:diagram} ($1$, $2$b, $2$c, $3$a, ...).

Firstly we define the bare susceptibility as
\begin{align*}
\chi^{0}_{o_1o_2o_3o_4}(\bm{q},\tau)=\frac{1}{N}\sum_{\bm{k}\bm{k}'}\langle T_{\tau} &c^{\dag}_{l_3,\bm{k}+\bm{q},\sigma}(\tau)c_{l_4,\bm{k},\sigma}(\tau) \ \times  \\ &c^{\dag}_{l_2,\bm{k}'-\bm{q},\sigma}(0)c_{l_1,\bm{k}',\sigma}(0) \rangle_0 ,
\end{align*}
where $o_i$ are the orbital indices. The bare susceptibility in momentum-frequency space is
\begin{widetext}
	\begin{align*}
	&\chi^0_{o_1o_2o_3o_4}(\bm{q},i\omega_n)=-\frac{1}{N}\sum_{\mu\nu}\sum_{\bm{k}}a^{o_4}_\mu(\bm{k})a^{o_2*}_{\mu}(\bm{k}) a^{o_1}_\nu(\bm{k}+\bm{q}) a^{o_3*}_{\nu}(\bm{k}+\bm{q})
	\frac{n_F(E_{\mu}(\bm{k}))-n_F(E_{\nu}(\bm{k}+\bm{q}))}{i\omega_n+E_{\mu}(\bm{k})-E_{\nu}(\bm{k}+\bm{q})}\ .
	\end{align*}
\end{widetext}
where $\mu/\nu$ is the band index. $n_F(\epsilon)$ is the Fermi distribution function, $a^{o_i}_\mu(\bm{k})$ is the $o_i$-th component of the eigenvector for band $\mu$ resulting from the diagonalization of the single-particle Hamiltonian $H_0$, and $E_{\mu}(\bf{k})$ is the eigenvalue of band $\mu$. The intrinsic spin fluctuations are characterized by the susceptibility. The interacting spin susceptibility and charge susceptibility in RPA level are given by,
\begin{align*}
\chi^{RPA}_1(\bm{q})&=[1-\chi_0(\bm{q})U^s]^{-1}\chi_0(\bm{q})\ ,\\[5pt]
\chi^{RPA}_0(\bm{q})&=[1+\chi_0(\bm{q})U^c]^{-1}\chi_0(\bm{q}) .
\end{align*}
Here $U^s$, $U^c$ are the interaction matrices, which are given by
\begin{align*}
\bar{U}^s_{o_1o_2o_3o_4}&=
\begin{cases}
U \phantom{~+2J+} & o_1=o_2=o_3=o_4,\\
V   & o_1=o_3\neq o_2=o_4,\\
J   & o_1=o_2\neq o_3=o_4,\\
J'   & o_1=o_4\neq o_2=o_3,\\
\end{cases}\\
\bar{U}^c_{o_1o_2o_3o_4}&=
\begin{cases}
U   & o_1=o_2=o_3=o_4,\\
-V+2J   & o_1=o_3\neq o_2=o_4,\\
2\,V-J   & o_1=o_2\neq o_3=o_4,\\
J'   & o_1=o_4\neq o_2=o_3,\\
\end{cases}
\end{align*}
where we used the notation for the Kanamori interaction parameters introduced in the main text.

The effective interaction obtained in the RPA approximation is given by
\begin{align*} V_{\rm eff}=\sum_{ij,\textbf{k}\textbf{k}'}\Gamma_{ij}(\textbf{k},\textbf{k}')c^{\dag}_{i\textbf{k}\uparrow}c^{\dag}_{i-\textbf{k}\downarrow}c_{j-\textbf{k}'\downarrow}c_{j\textbf{k}'\uparrow}
\end{align*}
where the momenta $\textbf{k}$ and $\textbf{k}'$ are restricted to  different FS $C_i$ with $\textbf{k}\in C_i$ and $\textbf{k}'\in C_j$, and $\Gamma_{ij}(\textbf{k},\textbf{k}')$ is the pairing scattering vertex. The pairing vertex can be obtained by projecting the pairing vertex in orbital space onto Fermi surfaces,
\begin{align*}
&\Gamma_{ij}(\textbf{k},\textbf{k}')=  \sum_{\substack{o_1,o_2 \\ o_3,o_4}}a^{o_2*}_{v_i}(\textbf{k}) a^{o_3*}_{v_i}(-\textbf{k}) \ \times \\
&\qquad\qquad \text{Re}[\Gamma_{o_1 o_2 o_3 o_4}(\textbf{k},\textbf{k}',\omega=0)] a^{o_1}_{v_j}(\textbf{k}') a^{o_4}_{v_j}(-\textbf{k}').
\end{align*}
The orbital vertex function $\Gamma_{o_1 o_2 o_3 o_4}$ for the singlet channel and triplet channel in the fluctuation exchange formulation\cite{TTakimoto-PRB-69-104504,KKubo-PRB-75-224509,SGraser-NJP-11-025016,AFKemper-NJP-12-073030} are given by
\begin{widetext}
	\begin{align*}
	\Gamma^S_{o_1 o_2 o_3 o_4}(\textbf{k}, \textbf{k}',\omega)&=[\frac{3}{2}\bar{U}^s \chi^{RPA}_1(\textbf{k}-\textbf{k}',\omega)\bar{U}^s + \frac{1}{2}\bar{U}^s   -\frac{1}{2}\bar{U}^c\chi^{RPA}_0(\textbf{k}-\textbf{k}',\omega)\bar{U}^c+\frac{1}{2}\bar{U}^c]_{o_1 o_2 o_3 o4},\\[10pt]
	\Gamma^T_{o_1 o_2 o_3 o_4}(\textbf{k}, \textbf{k}',\omega)&=[-\frac{1}{2}\bar{U}^s \chi^{RPA}_1(\textbf{k}-\textbf{k}',\omega)\bar{U}^s+\frac{1}{2}\bar{U}^s  -\frac{1}{2}\bar{U}^c\chi^{RPA}_0(\textbf{k}-\textbf{k}',\omega)\bar{U}^c+\frac{1}{2}\bar{U}^c]_{o_1 o_2 o_3 o_4},
	\end{align*}
\end{widetext}
where $\bar{U}^{s/c}=U^{s/c}(\bm{k}-\bm{k}')$.
$\chi^{RPA}_0$ describes the charge fluctuation contribution and $\chi^{RPA}_1$ the spin fluctuation contribution. For a given gap function $g(\textbf{k})$, the pairing strength functional is
\begin{align*}
\lambda[g(\textbf{k})]=-\frac{\sum_{ij}\oint_{C_i} \frac{dk_{\|}}{v_F(\textbf{k})} \oint_{C_j} \frac{dk'_{\|}}{v_F(\textbf{k}')} g(\textbf{k})\Gamma_{ij}(\textbf{k},\textbf{k}') g(\textbf{k}')} {4\pi^2\sum_i\oint_{C_i} \frac{dk_{\|}}{v_F(\textbf{k})} [g(\textbf{k})]^2 },
\end{align*}
where $v_F(\textbf{k})=|\triangledown_{\textbf{k}}E_i(\textbf{k})|$ is the Fermi velocity on a given Fermi surface sheet $C_i$. From the stationary condition we find the following eigenvalue problem:
\begin{align*}
-\sum_{j} \oint_{C_j} \frac{dk'_{\|}}{4\pi^2v_F(\textbf{k}')} \Gamma_{ij}(\textbf{k},\textbf{k}') g_{\alpha}(\textbf{k}')=\lambda_{\alpha}g_{\alpha}(\textbf{k}),
\end{align*}
where the interaction $\Gamma_{ij}$ is the symmetric (antisymmetric) part of the full interaction in the singlet (triplet) channel. The leading eigenfunction $g_{\alpha}(\bf{k})$ and eigenvalue $\lambda_{\alpha}$ are obtained from the above equation. The obtained gap function should have the symmetry of one of the irreducible representations for the corresponding point group.

\begin{figure}[t!]
	\includegraphics[width=1\columnwidth]{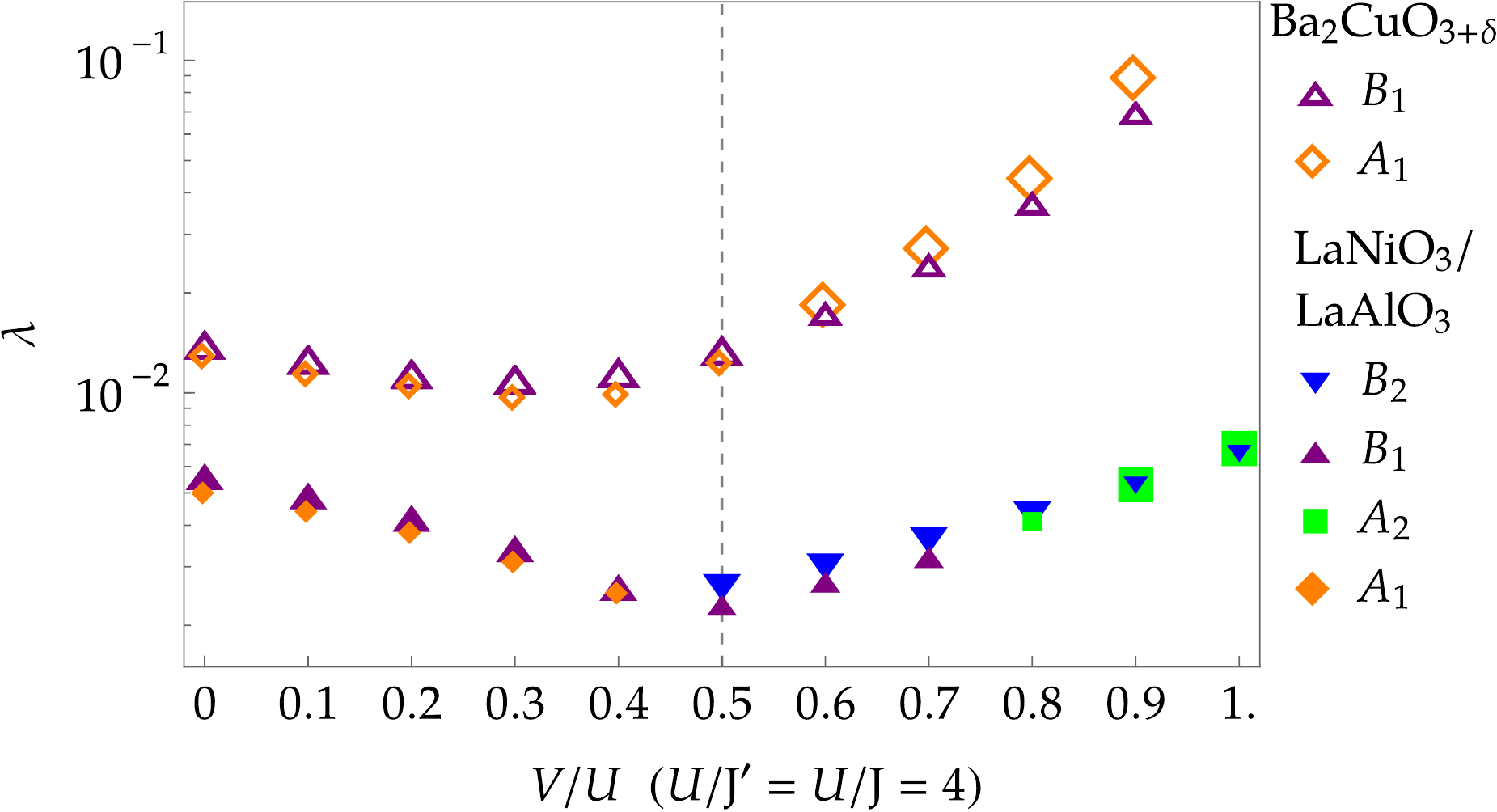}
	\caption{\label{fig:PhasediagrammRPA} Leading and subleading eigenvalues $\lambda$ within RPA and symmetry character of the superconducting order parameter of the effective four point RPA vertex $\Gamma$ for LNO/LAO (solid markers) and for BCO (unfilled markers). The interaction is given by $U=4J=4J'=0.8\,\mbox{eV}$ and increasing $V$ and $T=0.02\,$eV. The dashed line indicates the the generic interaction ratio $U=2V=4J=4J'$. For BCO  $\Gamma$ diverges for $V/U=1$, signalling the onset of a spin density wave.}
\end{figure}

\textit{Results.}---We show the leading and first subleading eigenvalues calculated within the RPA-matrix formalism for \ce{LaNiO3} and \ce{Ba2CuO_{3+\delta}} in Fig.\,\ref{fig:PhasediagrammRPA}.
The eigenvalues of \ce{Ba2CuO_{3+\delta}} are for the whole parameter regime much larger than \ce{LaNiO3}, due to the overall better nesting. As expected, a large regime of interactions is favoring $d$-wave pairing ($B_1$ irrep), given the large signal in the bare susceptibility. For $V/U > 1$ the extended $s$-wave pairing ($A_1$ irrep), featuring nodes between the Fermi pockets, becomes leading.

In case of \ce{LaNiO3}, there is no clear peak in the bare susceptibility, further the inner Fermi pocket around the $\Gamma$-point is largely inert to superconductivity in a repulsive interaction environment, since it features no density fluctuations, is clean in orbital weight, and is in itself not nested. Due to the density distribution of the outer Fermi sheet around the $M$ point, which is enlarged at the edges of the BZ, the resulting $B_1$ symmetry character is reminiscent of the one band cuprate models. Increasing inter-orbital interactions enlarges the coupling between the parallel lines on the outer Fermi pocket. Hence, additional nodal lines are required, resulting in a $B_2$ or $A_2$ symmetry character. Note that there is, similarly to the results of fRG, a pronounced dip in the eigenvalues of  LNO/LAO starting at the pure intra-orbital case and increasing inter-orbital interactions.

Using the eigenvalues from the RPA calculations, we can estimate the transition temperature as $T_c\approx\hbar\omega_Se^{-1/|\lambda_{\rm min}|}$, where $\hbar\omega_S$ is the typical energy scale for spin fluctuations in systems. The T$_c$ of BCO is about 70 K in experiments. To reproduce the experimental value approximately, we take $U=1$ eV, $J/U=0.25$, $\hbar\omega_S=100$ meV and the temperature parameter $T=0.03$ eV in the Fermi distribution function and then get the dominant pairing state with $\lambda_{\rm min}=-0.448$ for BCO, leading to the $T_c$ of the order of 100 K. With the same parameters in  LNO/LAO, the dominant pairing eigenvalue is 0.005, much smaller than the case of BCO.

\subsection{Weak coupling Renormalization Group - Kohn-Luttinger method}

\textit{Method.}---Here, we give a short description of the (KL) method; a detailed discussion can be found in various works \cite{raghu_superconductivity_2010,raghu_superconductivity_2011,raghu_effects_2012,cho_band_2013,wolf_unconventional_2018,wolf-20arXiv2004.12624}. Note that one of the most important characteristics of the method is that it becomes asymptotically exact in the limit of infinitesimal coupling, $U\rightarrow 0$.

For the KL analysis, we consider all terms in the perturbative expansion of the effective two-particle-vertex up to second order in the interaction, $U$, which is given by the first four diagrams in Fig.\,\ref{Fig:App:diagram}, \ie diagrams 1, 2a, 2b, and 2c.
In contrast to fRG, where higher order diagrams are considered gradually during the flow of the effective interaction parameters, we cut off all diagrams that are of higher than second order in $U$. In RPA, on the contrary, we sum over a geometric series of only ladder and bubble diagrams, \ie up to infinite order in $U$.

\begin{figure}[b!]
	\includegraphics[width=0.9\columnwidth]{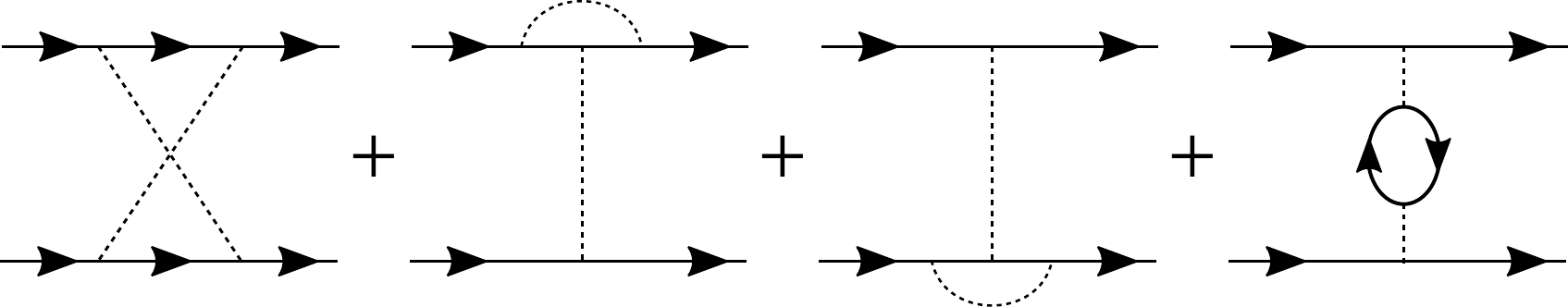}
	\caption{Spinless particle-hole diagrams of the weak coupling expansion of the effective two-particle-vertex in second order in the interaction $U$. Solid lines represent free particle propagators and dashed lines interactions $U$.}
	\label{Fig:App:WCRG_diagram}
\end{figure}

The RG flow equation is given by \cite{shankar_renormalization-group_1994,raghu_superconductivity_2010,wolf_unconventional_2018}
\begin{equation}
\label{Eq:App:WCRG_flow}
\frac{\partial \Gamma(k_{2},k_{1})}{\partial \ln(\Omega_0/\Omega)}=-\int_{\rm FS}{\rm d}k_{3}\,\Gamma(k_{2},k_{3})\Gamma(k_{3},k_{1}),
\end{equation}
where $\Gamma$ denotes the effective two-particle-vertex, and $k_{1}$ and $k_{2}$ denote the momenta of the incoming and outgoing pairs of electrons with zero total momentum, respectively. The right hand side is integrated over all Fermi surfaces and we used the short notation $k_{i}=(n_{i},\vec{k}_{i})$ with the band index $n_{i}$. $\Omega_0$ denotes the initial infrared cutoff and $\Omega$ a lowered cutoff. The initial cutoff is chosen in the range
\begin{equation}
\label{eq:cutoff_range}
U^{2}/W\gg\Omega_{0}\gg We^{-1/\rho|U|},
\end{equation}
where $W$ is the bandwidth.
The lower bound ensures that the bare interactions are renormalized due to many-body effects, whereas the upper bound limits all involved modes in the effective interaction to a narrow window around the Fermi surface\,\cite{raghu_superconductivity_2010}.

In the weak coupling limit we can bypass the procedure of the RG flow, \ie lowering the cutoff $\Omega$ in consecutive steps, and obtain the final $\Gamma$ directly, where Eq.\,\eqref{eq:cutoff_range} ensures that the final result does not depend on the initial cutoff, $\Omega_{0}$\,\cite{raghu_superconductivity_2010}. All terms which contribute in this manner are the diagrams 2b and 2c in Fig.\,\ref{Fig:App:diagram}, where we use their spinless variants which are given by the four diagrams shown in Fig.\,\ref{Fig:App:WCRG_diagram} (the first order in $U$ diagram just suppresses the plain $s$-wave solution and can thus be neglected).

%
\begin{figure}[t]
	\includegraphics[width=1\columnwidth]{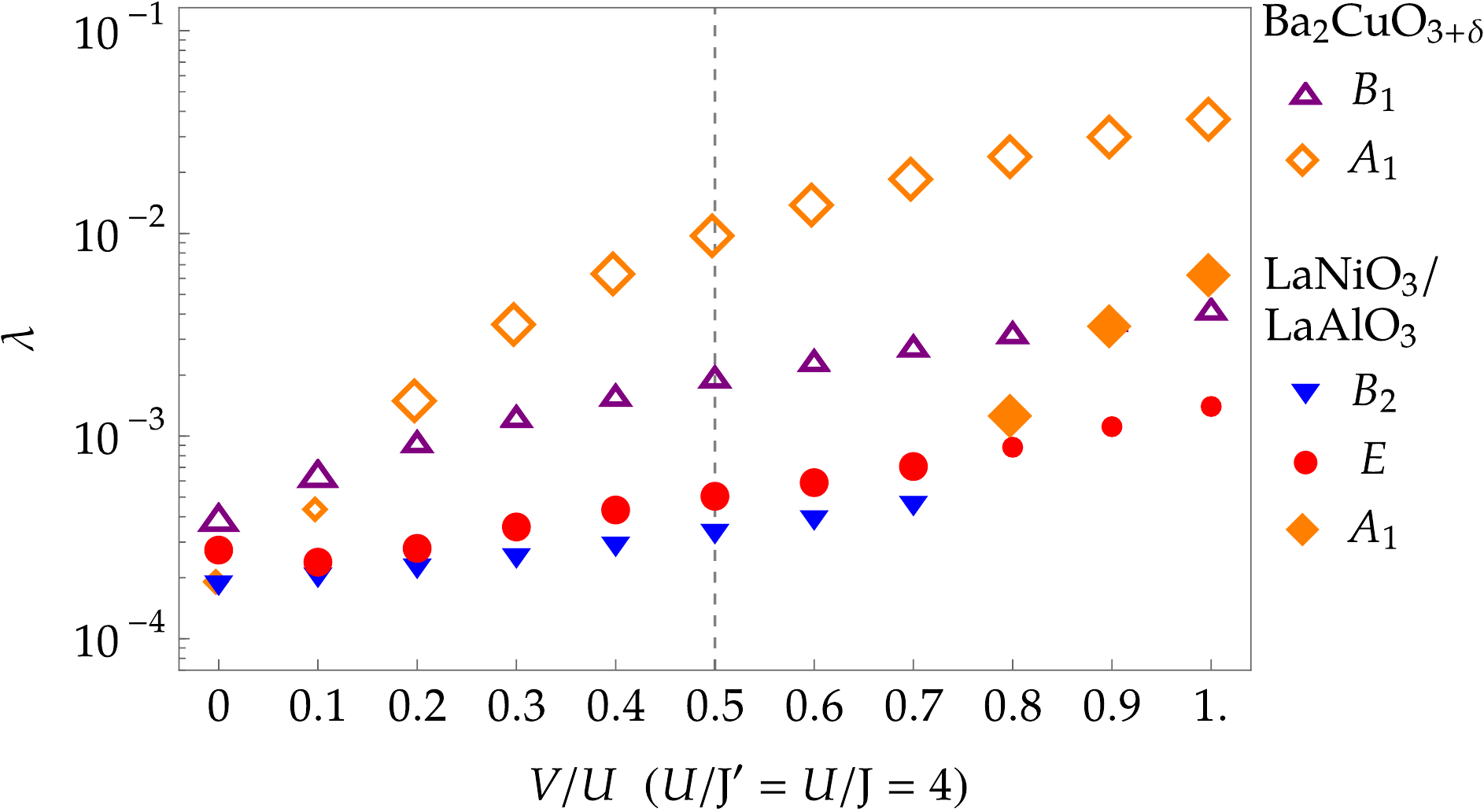}
	\caption{\label{fig:Phasediagramm_wcRG} KL Results: leading and subleading Eigenvalues $\lambda$ and symmetry character of leading superconducting order of effective four point vertex $\Gamma$ forLNO/LAO (solid markers) and BCO (unfilled markers). The interaction is given by $U=4J=4J'$ and increasing $V$. The dashed line indicates the the generic interaction ratio $U=2V=4J=4J'$.}
\end{figure}

In Eq.\,\eqref{Eq:App:WCRG_flow} $\Gamma$ is scaled such that its eigenvalues $\lambda_i$, independently, fulfil
\begin{equation}
\frac{\partial \lambda_i}{\partial\ln(\Omega_{0}/\Omega)}=-\lambda_i^{2}.
\end{equation}
Hence, finding the leading superconducting instability simplifies to an eigenvalue problem for $\Gamma$. The effective interaction, $V_{\rm eff}=\lambda_{\rm min}/\rho$, of the leading superconducting instability is then obtained from the most negative eigenvalue, $\lambda_{\rm min}$ ($\rho$ is the total density of states at the Fermi level), and the relation to the critical temperature is given by
\begin{equation*}
T_{c}\sim e^{-1/\rho |V_{\rm eff}|}=e^{-1/|\lambda_{\rm min}|}.
\end{equation*}
The corresponding eigenfunction, $\psi_{\rm min}$, yields the formfactor of the order parameter, which can be classified by the irreducible representations of the symmetry group of the crystal lattice (analogous to the fRG as shown in the main paper).

\textit{Results.}---The eigenvalues of the leading superconducting instabilities for each irreducible representation obtained from the KL formalism are shown in Fig.\,\ref{fig:Phasediagramm_wcRG}. Here we plot the eigenvalue $\lambda_{\rm min}$ as a function of the interorbital interaction, $V$, while keeping the other interactions at $U=4J=4J'$.

Firstly, we note that $|\lambda_{\rm min}|$ for Ba$_{2}$CuO$_{3+\delta}$ is larger than for LaNiO$_{3}$ over the whole parameter range. For both systems, an extended $s$-wave solution ($A_1$ irrep) becomes dominant when $V/U$ approaches $1$, in agreement with the fRG and RPA results.

For Ba$_{2}$CuO$_{3+\delta}$, this s-wave ($A_{1}$ irrep) is also the leading instability for an extended range around $U=2V$. For small $V$, there is a transition to a $d_{x^2-y^2}$-wave state ($B_1$ irrep). 

Note that for fRG, for which we used the highest bare interaction the regime of the $A_{1}$ irrep in the $V/U$ phase diagram is the smallest compared to the other methods (transition at $V/U\gtrsim 0.9$). In RPA this regime is enlarged (transition at $V/U\approx 0.5 \,...\, 0.6$) and the KL analysis, where U is infinitesimal yields the largest domain with a leading s-wave (transition at $V/U\approx 0.2  \,...\, 0.3$)

For LaNiO$_{3}$ we find a $p$-wave ($E$ irrep) solution for most of the shown parameter range, until the transition towards extended $s$-wave at large values for $V$.

As mentioned in the main paper, due to the asymptotic character, \ie the vanishing interacting strength, the KL analysis is extremely sensitive to FS fluctuations; thus differences for the Nickelates compared to RPA and fRG are expected.

\bibliography{bib,SM_bib}

\end{document}